\documentclass[twocolumn]{article}
\usepackage{lmodern}
\usepackage{amssymb,amsmath}
\usepackage{ifxetex,ifluatex}
\usepackage{fixltx2e} 
\ifnum 0\ifxetex 1\fi\ifluatex 1\fi=0 
  \usepackage[T1]{fontenc}
  \usepackage[utf8]{inputenc}
\else 
  \ifxetex
    \usepackage{mathspec}
  \else
    \usepackage{fontspec}
  \fi
  \defaultfontfeatures{Ligatures=TeX,Scale=MatchLowercase}
\fi
\IfFileExists{upquote.sty}{\usepackage{upquote}}{}
\IfFileExists{microtype.sty}{%
\usepackage{microtype}
\UseMicrotypeSet[protrusion]{basicmath} 
}{}
\usepackage[margin=1in]{geometry}
\usepackage{hyperref}
\hypersetup{unicode=true,
            pdftitle={How to Read a Research Compendium},
            pdfauthor={Daniel Nüst, Institute for Geoinformatics, University~of~Münster, Münster,~Germany (daniel.nuest@uni-muenster.de); Carl Boettiger, Department of Environmental Science, Policy and Management, University~of~California,~Berkeley, Berkeley,~California, United~States (cboettig@gmail.com); Ben Marwick, Department of Anthropology, University~of~Washington, Seattle,~Washington, United~States (bmarwick@uw.edu)},
            pdfborder={0 0 0},
            breaklinks=true}
\usepackage{graphicx,grffile}
\makeatletter
\def\maxwidth{\ifdim\Gin@nat@width>\linewidth\linewidth\else\Gin@nat@width\fi}
\def\maxheight{\ifdim\Gin@nat@height>\textheight\textheight\else\Gin@nat@height\fi}
\makeatother
\setkeys{Gin}{width=\maxwidth,height=\maxheight,keepaspectratio}
\IfFileExists{parskip.sty}{%
\usepackage{parskip}
}{
\setlength{\parindent}{0pt}
\setlength{\parskip}{6pt plus 2pt minus 1pt}
}
\providecommand{\tightlist}{%
  \setlength{\itemsep}{0pt}\setlength{\parskip}{0pt}}
\ifx\paragraph\undefined\else
\let\oldparagraph\paragraph
\renewcommand{\paragraph}[1]{\oldparagraph{#1}\mbox{}}
\fi
\ifx\subparagraph\undefined\else
\let\oldsubparagraph\subparagraph
\renewcommand{\subparagraph}[1]{\oldsubparagraph{#1}\mbox{}}
\fi

\let\rmarkdownfootnote\footnote%
\def\footnote{\protect\rmarkdownfootnote}

\usepackage{titling}

  \title{How to Read a Research Compendium}
  \pretitle{\vspace{\droptitle}\centering\huge}
  \posttitle{\par}
  \author{Daniel Nüst, Institute for Geoinformatics, University~of~Münster,
Münster,~Germany
(\href{mailto:daniel.nuest@uni-muenster.de}{\nolinkurl{daniel.nuest@uni-muenster.de}}) \\ Carl Boettiger, Department of Environmental Science, Policy and
Management, University~of~California,~Berkeley, Berkeley,~California,
United~States
(\href{mailto:cboettig@gmail.com}{\nolinkurl{cboettig@gmail.com}}) \\ Ben Marwick, Department of Anthropology, University~of~Washington,
Seattle,~Washington, United~States
(\href{mailto:bmarwick@uw.edu}{\nolinkurl{bmarwick@uw.edu}})}
  \preauthor{\centering\large\emph}
  \postauthor{\par}
  \date{}
  \predate{}\postdate{}

 \usepackage{url} \usepackage{breakurl}
\PassOptionsToPackage{hyperindex,breaklinks}{hyperref}

\begin{document}
\maketitle
\begin{abstract}
Researchers spend a great deal of time reading research papers. Keshav
(2012) provides a three-pass method to researchers to improve their
reading skills. This article extends Keshav's method for reading a
research compendium. Research compendia are an increasingly used form of
publication, which packages not only the research paper's text and
figures, but also all data and software for better reproducibility. We
introduce the existing conventions for research compendia and suggest
how to utilise their shared properties in a structured reading process.
Unlike the original, this article is not build upon a long history but
intends to provide guidance at the outset of an emerging practice.
\end{abstract}

\hypertarget{introduction}{%
\section*{1. Introduction}\label{introduction}}
\addcontentsline{toc}{section}{1. Introduction}

\hypertarget{motivation}{%
\subsection*{1.1 Motivation}\label{motivation}}
\addcontentsline{toc}{subsection}{1.1 Motivation}

Research compendia are an increasingly used form of publication and
scholarly communication. They comprise not only the research paper's
text and figures, but also all data and software used to conduct the
computational workflow and create all outputs. They provide a lot of
added value by revealing more of the research process to readers, but,
if not done well, they can increase the difficulty of understanding the
research. To help readers better understand how to read a research
compendium, we extends Keshav's three-pass method targeted at improving
skills for reading a research paper (Keshav 2007) with additional steps
relevant to a research compendium's content.

Unlike the first version of the original (Keshav 2007), we cannot draw
from a long history of experience, because until recently research
compendia have been relatively rare. Our intention here is to provide
guidance at the outset of an emerging practice to both readers and
authors of research compendia to help them understand each others'
perspectives and needs and improve their communication. Authors can use
this guide to improve their research compendium's structure and content
by better anticipating their readers' needs. They should not be held
back by unwarranted concerns, like providing support (Barnes 2010).
Readers can avoid the trap of falling too deep into technological
challenges by an iterative approach to reading and using that gives
attention to the scientific issues. Ultimately research compendia can
enhance and deepen the reading experience, if done right. Keshav's
following introduction applies directly to research compendia:

\begin{quote}
\emph{Researchers must read papers for several reasons:} \emph{to review
them for a conference or a class, to keep current in their field, or for
a literature survey of a new field.} \emph{A typical researcher will
likely spend hundreds of hours every year reading papers.}

\emph{Learning to efficiently read a paper is a critical but rarely
taught skill.} \emph{Beginning graduate students, therefore, must learn
on their own using trial and error.} \emph{Students waste much effort in
the process and are frequently driven to frustration.}

\emph{For many years I have used a simple `three-pass' approach to
prevent me from drowning in the details of a paper before getting a
bird's-eye-view.} \emph{It allows me to estimate the amount of time
required to review a set of papers.} \emph{Moreover, I can adjust the
depth of paper evaluation depending on my needs and how much time I
have.} \emph{This paper describes the approach and its use in doing a
literature survey.} (Keshav 2016)
\end{quote}

The additions made in this work to accommodate for the content in a
research compendium are quite extensive. This stems from the complexity
that an interactive compendium has compared to a classic static
``paper'', because a research compendium goes well beyond the ``mere
advertising of the scholarship'' (Claerbout 1994). We see the breadth of
additions as a sign of potential, namely for unprecedented transparency,
openness, and collaboration.

\hypertarget{structure}{%
\subsubsection*{1.2 Structure}\label{structure}}
\addcontentsline{toc}{subsubsection}{1.2 Structure}

In the remainder of this paper, the excellent original work is taken
over completely. It is set in \emph{italic font} based on the most
recent online version: Keshav (2016). The term ``paper'' was not
replaced with ``research compendium'' for better readability.

First we briefly introduce research compendia and existing conventions.
We further list relevant resources for authors related to research
compendia. Then, matching the original paper's section numbering,
Sections 2 extends the ``Three-pass Approach'' to include research
compendium features in the reading process. Section 3 extends ``Doing a
Literature Survey'' with aspects relevant reviewing many research
compendia.

\hypertarget{research-compendia}{%
\subsubsection*{1.3 Research compendia}\label{research-compendia}}
\addcontentsline{toc}{subsubsection}{1.3 Research compendia}

The term \emph{research compendium} was coined by Gentleman and Lang
(2007) who ``introduce{[}d{]} the concept of a compendium as both a
container for the different elements that make up the document and its
computations (i.e.~text, code, data,\ldots{}), and as a means for
distributing, managing and updating the collection.'' According to
Marwick, Boettiger, and Mullen (2018) it provides ``a standard and
easily recognisable way for organising the digital materials of a
research project to enable other researchers to inspect, reproduce, and
extend the research''. This standard may differ between scientific
domains, yet the intentions and benefits are the same. Research
compendia are practised Open Science culture and as such improve
transparency (Nosek et al. 2015), ``make more published research true''
(Ioannidis 2014), and enable enhanced review and publication workflows
(Nüst et al. 2017). They answer readers' needs to understand complex
analyses through inspection and manipulation (Konkol and Kray 2018) and
enable other researchers to reproduce and extend the research (Marwick,
Boettiger, and Mullen 2018). Research compendia improve citations since
code and data are openly available (Vandevalle 2012). Ultimately, their
goal is to improve reproducibility (see Barba (2018) for definitions of
terms) in the light of claims of a ``reproducibility crisis'' in several
fields. Infrastructures to support the creation, scientific publication,
inspection, and collaboration based on research compendia are an active
field of research, but none of which have been widely deployed yet (Nüst
et al. (2017); Brinckman et al. (2018); Stodden, Miguez, and Seiler
(2015); Kluyver et al. (2016); Green and Clyburne-Sherin (2018)).

As this article is focused on providing hands-on guidance on using, and
to some extend also creating, research compendia, we refer the reader to
the references for more specific details. For the remainder of this
work, we assume a minimal view of a research compendium suitable for
\emph{readers} who examine a research compendium directly. A research
compendium has three integral parts: text, code, and data. Text can be
instructions, software documentation, or a full manuscript with figures.
Code can be scripts, software packages, specifications of dependencies
and computational environments, or even virtual machines. Data can be
just about anything, but probably comprises plain text or binary files
that are used as input to the workflow, and produced as output from
executing the workflow.

For \emph{authors}, there is a wealth of generic recommendations guiding
researchers in creating open research (software), for example Sandve et
al. (2013), Taschuk and Wilson (2017), Prlić and Procter (2012), Stodden
and Miguez (2014), and Wilson et al. (2017). When a research compendium
is published, one can assume the authors have the intention to help the
reader understanding the work and accepts there are ``no excuses'' to
not publishing your code (Barnes 2010). Authors may attempt to reach the
ideals of having one ``main'' file that can be executed with
``one-click'' (Pebesma 2013), of enabling re-use with proper licensing
(Stodden 2009), and of interweaving code and text following the literate
programming paradigm (Knuth 1984).

The following conventions are specifically for research compendia:

\begin{itemize}
\tightlist
\item
  Marwick, Boettiger, and Mullen (2018) and ROpenSci community's
  \texttt{rrrpkg} (\url{https://github.com/ropensci/rrrpkg}) discuss the
  standards and tooling of the R programming language and software
  engineering tools for a variety of disciplines with real-world
  examples, including several templates
\item
  Jimenez et al. (2017) apply software engineering best pratices from
  the Open Source software domain to research (see also
  \url{http://falsifiable.us/}).
\item
  Konkol, Kray, and Pfeiffer (2018) derive recommendations for authors
  from issues encountered reproducing research compendia in geosciences
\item
  Gentleman and Lang (2007) recommend using programming languages'
  packaging mechanisms for research compendia, more specifically R and
  Python packages
\item
  Chirigati et al. (2016) describe the tool \texttt{ReproZip}
  (\url{https://reprozip.org}) to support capture and reproduction of a
  research compendium
\end{itemize}

\hypertarget{the-three-pass-approach}{%
\section*{2. The three-pass approach}\label{the-three-pass-approach}}
\addcontentsline{toc}{section}{2. The three-pass approach}

\begin{quote}
\emph{The key idea is that you should read the paper in up to three
passes, instead of starting at the beginning and plowing your way to the
end.} \emph{Each pass accomplishes specific goals and builds upon the
previous pass:} \emph{The first pass gives you a general idea about the
paper.} \emph{The second pass lets you grasp the paper's content, but
not its details.} \emph{The third pass helps you understand the paper in
depth.} (Keshav 2016)
\end{quote}

\hypertarget{the-first-pass}{%
\subsection*{2.1 The first pass}\label{the-first-pass}}
\addcontentsline{toc}{subsection}{2.1 The first pass}

\begin{quote}
\emph{The first pass is a quick scan to get a bird's-eye view of the
paper.} \emph{You can also decide whether you need to do any more
passes.} \emph{This pass should take about five to ten minutes and
consists of the following steps:} (Keshav 2016)
\end{quote}

\begin{enumerate}
\def\labelenumi{\arabic{enumi}.}
\tightlist
\item
  \emph{Carefully read the title, abstract, and introduction}
\item
  \emph{Read the section and sub-section headings, but ignore everything
  else}
\item
  \emph{Glance at the mathematical content (if any) to determine the
  underlying theoretical foundations}
\item
  \emph{Read the conclusions}
\item
  \emph{Glance over the references, mentally ticking off the ones you've
  already read}
\item
  Glance over the text looking for (a) URLs and \texttt{formatted}
  \textbf{names} referencing software and data products or repositories
  not yet mentioned in the sections read so far, mentally ticking off
  the ones you've heard about or used, and (b) tables or figures
  describing computational environments, deployments, or execution
  statistics
\end{enumerate}

\emph{At the end of the first pass, you should be able to answer the}
seven \emph{Cs:}

\begin{enumerate}
\def\labelenumi{\arabic{enumi}.}
\tightlist
\item
  \emph{Category: What type of paper is this? A measurement paper? An
  analysis of an existing system? A description of a research
  prototype?}
\item
  \emph{Context: Which other papers is it related to? Which theoretical
  bases were used to analyze the problem?}
\item
  \emph{Correctness: Do the assumptions appear to be valid?}
\item
  \emph{Contributions: What are the paper's main contributions?}
\item
  \emph{Clarity: Is the paper well written?}
\item
  Construction: What are the building blocks of the analysis workflow
  and how accessible are they (data set(s), programming language(s),
  tools, algorithms, scripts)? Under what licenses are code and data
  published?
\item
  Complexity: What is the scale of the analysis (e.g.~HPC, required
  OS/cores/memory, typical execution time, data size) and the software
  (number of dependencies and is installation possible with dependency
  management tools)?
\end{enumerate}

\begin{quote}
\emph{Using this information, you may choose not to read further (and
not print it out, thus saving trees).} \emph{This could be because the
paper doesn't interest you, or you don't know enough about the area to
understand the paper, or that the authors make invalid assumptions.}
(Keshav 2016)
\end{quote}

You may also choose not to pursue the parts of the research compendium
further, i.e.~not running the workflow or looking at data or code, thus
saving resources. Reasons to not read further that relate specifically
to code and data may be that you don't have the expertise or access to
resources to re-use the data and code.

\begin{quote}
\emph{The first pass is adequate for papers that aren't in your research
area, but may someday prove relevant.} (Keshav 2016)
\end{quote}

This first pass suits research compendia comprising potentially
re-usable components, like workflows or algorithms using data sets or
generic software that are directly transferable to your field of
research. After the first pass, you should be able to judge if the
software is useful, if it works.

\begin{quote}
\emph{Incidentally, when you write a paper, you can expect most
reviewers (and readers) to make only one pass over it.} \emph{Take care
to choose coherent section and sub-section titles and to write concise
and comprehensive abstracts.} \emph{If a reviewer cannot understand the
gist after one pass, the paper will likely be rejected; if a reader
cannot understand the highlights of the paper after five minutes, the
paper will likely never be read.} \emph{For these reasons, a `graphical
abstract' that summarizes a paper with a single well-chosen figure is an
excellent idea and can be increasingly found in scientific journals.}
(Keshav 2016)
\end{quote}

When you write a paper, take care to add instructions on how a reader
can reproduce your work and provide all required parts, i.e.~publish a
research compendium. The instructions should start with a ``blank''
system and be specific, i.e.~ready for copy \& paste, including expected
or experienced execution times and resources. Such instructions give
readers a good idea about what is needed to recreate your environment
and execute the analysis If your work requires specialised or bespoke
hardware (HPC, specific GPUs), consider creating an exemplary, reduced
analysis that runs in regular environments.

Also ensure your code and data are properly deposited, citable and
licensed. If you don't do this, these core parts of your work will
likely never be properly evaluated or re-used. See the section
``Research Compendia'', above, for recommendations and further reading
on how to make your reviewers' and readers' lives easier.

\hypertarget{the-second-pass}{%
\subsection*{2.2 The second pass}\label{the-second-pass}}
\addcontentsline{toc}{subsection}{2.2 The second pass}

\begin{quote}
\emph{In the second pass, read the paper with greater care, but ignore
details such as proofs.} \emph{It helps to jot down the key points, or
to make comments in the margins, as you read.} \emph{Dominik Grusemann
from Uni Augsburg suggests that you ``note down terms you didn't
understand, or questions you may want to ask the author.''} \emph{If you
are acting as a paper referee, these comments will help you when you are
writing your review, and to back up your review during the program
committee meeting.} (Keshav 2016)
\end{quote}

\begin{enumerate}
\def\labelenumi{\arabic{enumi}.}
\tightlist
\item
  \emph{Look carefully at the figures, diagrams and other illustrations
  in the paper. Pay special attention to graphs. Are the axes properly
  labelled?} \emph{Are results shown with error bars, so that
  conclusions are statistically significant? Common mistakes like these
  will separate rushed, shoddy work from the truly excellent.} (Keshav
  2016)
\item
  \emph{Remember to mark relevant unread references for further reading
  (this is a good way to learn more about the background of the paper).}
  (Keshav 2016)
\item
  Skim over data and source code files without opening them. Are they
  reasonably named (Bryan 2015)? Do they follow a well-defined structure
  (e.g.~a Python package or a research compendium convention)? Is there
  a README file and/or structured documentation for functionalities?
\item
  Visit the online source code repository, if available. Is it
  established and well maintained, or orphaned? Is there only one author
  or are there contributors? How responsive are they to issues? Does the
  repository have signs of public recognition (i.e.~GitHub ``stars'' and
  ``forks'')? Are there regular releases, using semantic versioning?
\item
  Follow the instructions to install the required software and execute
  the research compendium's workflow with the provided parameters and
  input or sample data. Note down errors or warnings but do not try to
  fix any but trivial or known problems (e.g.~fixing a path or
  installing an undocumented dependency).
\item
  Compare the outputs with the expected ones reported in the paper. Also
  check for differences in output figures: Do labels, legends etc. match
  those in the paper?
\end{enumerate}

Points 3 and 4 above hint at how to estimate the quality of a software,
but we recommend to be realistic as to what to expect and be careful not
to judge too fast. The software project you evaluate might be done by a
single researcher who is not a professional programmer, working under a
lot of pressure to write code for a single use case. In these situations
one might find low levels of code documentation, but further
documentation might be quickly provided by the authors once you as an
external user show interest. Also, no recent changes or releases at a
source code repository can also mean the software is stable and simply
works with no problems!

\begin{quote}
\emph{The second pass should take up to an hour for an experienced
reader.} (Keshav 2016)
\end{quote}

This does not include the computation time of workflows in a research
compendium. Use this time to complete first passes for one or several
other compendia. If the software used is familiar, you may attempt to
reduce the computation time by sub-setting data or simplifying the
workflow. As an author, consider adding a reduced example to your
research compendium for easier access by readers.

\begin{quote}
\emph{After this pass, you should be able to grasp the content of the
paper.} (Keshav 2016)
\end{quote}

You should have re-executed the provided workflow or understand why you
could not. You should be able to complete the second pass even if you
are unfamiliar with the actual language the software is written in or if
you are not a developer yourself. However we do recommend not to dive
too deep, i.e.~not going beyond the provided instructions for the
research compendium's workflow. At this stage, it is the author's
responsibility to guide you through their work.

Still, you may also face unsolvable problems, like access to specific
infrastructure. But if you encounter issues or have questions, you
should communicate these to the author, for example in the software's
public code repository, if available. It is important to do this
respectfully, and give the authors a chance to fix bugs or respond to
issues (Kahneman 2014). Also let the authors know if your reproduction
was successful, especially if you used a different operating system or
software version than reported.

At this point you should be able to judge whether the software works and
if it is sustainable. Based on this evaluation you can decide to re-use
parts of the analysis, i.e.~software, data, or method, for your own
work.

\begin{quote}
\emph{You should be able to summarize the main thrust of the paper, with
supporting evidence, to someone else.} \emph{This level of detail is
appropriate for a paper in which you are interested, but does not lie in
your research speciality.} \emph{Sometimes you won't understand a paper
even at the end of the second pass.} \emph{This may be because the
subject matter is new to you, with unfamiliar terminology and acronyms.}
\emph{Or the authors may use a proof or experimental technique that you
don't understand, so that the bulk of the paper is incomprehensible.}
\emph{The paper may be poorly written with unsubstantiated assertions
and numerous forward references.} (Keshav 2016)
\end{quote}

The research compendium may have incomplete documentation, rely on
unavailable software (e.g.~proprietary) or data (e.g.~sensitive), or
require infrastructure not available to you (e.g.~high-performance
computing, HPC). It may use a programming language or programming
paradigms unfamiliar to you.

\begin{quote}
\emph{Or it could just be that it's late at night and you're tired.}
\emph{You can now choose to: (a) set the paper aside, hoping you don't
need to understand the material to be successful in your career, (b)
return to the paper later, perhaps after reading background material or
(c) persevere and go on to the third pass.} (Keshav 2016)
\end{quote}

\hypertarget{the-third-pass}{%
\subsection*{2.3 The third pass}\label{the-third-pass}}
\addcontentsline{toc}{subsection}{2.3 The third pass}

\begin{quote}
\emph{To fully understand a paper, particularly if you are a reviewer,
requires a third pass.} \emph{The key to the third pass is to attempt to
virtually re-implement the paper: that is, making the same assumptions
as the authors, re-create the work.} \emph{By comparing this re-creation
with the actual paper, you can easily identify not only a paper's
innovations, but also its hidden failings and assumptions.} \emph{This
pass requires great attention to detail.} (Keshav 2016)
\end{quote}

If a best practice or established convention for structuring data and
code was followed, familiarise yourself with it now.

\begin{quote}
\emph{You should identify and challenge every assumption in every
statement.} \emph{Moreover, you should think about how you yourself
would present a particular idea.} \emph{This comparison of the actual
with the virtual lends a sharp insight into the proof and presentation
techniques in the paper and you can very likely add this to your
repertoire of tools.} (Keshav 2016)
\end{quote}

Take a close look at data, metadata, source code including the embedded
code comments, and further documentation. You now leave the realm of the
mere software user to the developer's perspective. This can be a time
consuming very close study of the materials. If data is not publicly
available, e.g.~because it contains information about human subjects,
decide if you have a reasonable request to contact the original authors
and ask for data access. Work though the examples and analysis scripts
included in the research compendium. Play close attention not only to
code, but also to code comments as they should include helpful
information. A good entry point for your code read may be a ``main''
script (if provided by the author), makefile, or literate programming
document (e.g.~an R Markdown file or Jupyter Notebook). If neither of
these are available, then start with the code creating the figures for
the article (e.g.~look for ``\texttt{plot}'' statements in the code) and
trace your way back through the code until you reach a statement where
the input data is read. Your impression of the code can help to inform
your impression of the article's quality.

If you did not succeed before but the work is relevant for you, spend
more time on getting the analysis to run on your computer. Do not
hesitate to contact the authors of the paper or authors of the software
for help, but follow common error reporting guidelines (e.g. Stack
Overflow (2018) or Tatham (n.d.)). For authors it is a great experience
to be contacted by an interested and respectful reader!

With regard to the analysis, you may re-implement core parts or the full
workflow with a different software. For example, using a tool you know
but which was not used in the research compendium. Does your code lead
to the same results, or does it give different ones? Can the differences
be explained or are they not significant? Note that such a replication
is of very high value for science and you should share your findings
with the research compendium's authors and also with the scientific
community. Depending on the efforts you put in, write a blog post or
even publish a replication research compendium for one or more evaluated
research compendia.

If a full replication is not feasible, explore the assumptions you
challenge with data and code. Play around with input parameters to get a
feel for the changing results. Create exploratory plots for the data as
if you would want to analyse it from scratch, without the knowledge of
the existing workflow. With your understanding of the code you can
extend the method to a new problem or apply it to a different dataset.
This deep evaluation of code and data increases your understanding of
the authors' reasoning and decisions, and may lead to new questions.

To make sure you can trace your own hands-on changes with the original
code and configuration. We recommend initiating a local git repository
when starting this pass. You can create branches for specific
explorations and easily reset to the original functional state.

\begin{quote}
\emph{During this pass, you should also jot down ideas for future work.}
\emph{This pass can take many hours for beginners and more than an hour
or two even for an experienced reader.} \emph{At the end of this pass,
you should be able to reconstruct the entire structure of the paper from
memory, as well as be able to identify its strong and weak points.}
\emph{In particular, you should be able to pinpoint implicit
assumptions, missing citations to relevant work, and potential issues
with experimental or analytical techniques.} (Keshav 2016)
\end{quote}

You should be able to come up with useful extensions of the used
software stack and be able to judge the transferability and reusability
of the analysis' building blocks. You should most certainly have
improved your programming skills by reading and evaluating other
people's code or even trying to extend or improve it.

\hypertarget{doing-a-literature-survey}{%
\section*{3. Doing a literature
survey}\label{doing-a-literature-survey}}
\addcontentsline{toc}{section}{3. Doing a literature survey}

\begin{quote}
\emph{Paper reading skills are put to the test in doing a literature
survey.} \emph{This will require you to read tens of papers, perhaps in
an unfamiliar field.} \emph{What papers should you read?} \emph{Here is
how you can use the three-pass approach to help.} \emph{First, use an
academic search engine such as Google Scholar or CiteSeer and some
well-chosen keywords to find three to five recent highly-cited papers in
the area.} (Keshav 2016)
\end{quote}

No search capability comparable to scientific articles exists for
research compendia, though you can of course use generic and academic
search engines. More and more journals encourage reproducible research
and software and data publication, so that extending your search regular
search with keywords such as ``reproduction'', ``reproducible'', ``open
data/software/code'' may improve your results.

In addition, you can search online platforms where research compendia
have been published and tagged as a research compendium
(\texttt{research-compendium}):

\begin{itemize}
\tightlist
\item
  GitHub~label: \url{https://github.com/topics/research-compendium}
\item
  Zenodo~community:
  \url{https://zenodo.org/communities/research-compendium}
\end{itemize}

There is no journal specifically for research compendia yet, but the
following ones feature reproducibility, computational studies, or
openness in a prominent way and can be a starting point for finding
research compendia, if they fit your topic:

\begin{itemize}
\tightlist
\item
  \emph{ReScience}: \url{https://rescience.github.io/}
\item
  \emph{Information~Systems} has a reproducibility editor and special
  track for invited reproducibility papers:
  \url{https://www.journals.elsevier.com/information-systems/}
\end{itemize}

A lateral approach takes advantage of the parts of a research
compendium. If you work with a specific software (tool, extension
package, library) or data, find out the recommended way to cite it (and
follow it yourself). Most scientific software provides this information
in their FAQ or might have a built-in function to generate a citation.
Scientific data is often accompanied by a ``data paper'' or published in
repositories with citeable identifiers. Then search for recent
publications which cite the referenced software or data.

\begin{quote}
\emph{Do one pass on each paper to get a sense of the work, then read
their related work sections.} \emph{You will find a thumbnail summary of
the recent work, and perhaps, if you are lucky, a pointer to a recent
survey paper.} \emph{If you can find such a survey, you are done.}
\emph{Read the survey, congratulating yourself on your good luck.}
\emph{Otherwise, in the second step, find shared citations and repeated
author names in the bibliography.} \emph{These are the key papers and
researchers in that area.}
\end{quote}

You can also find shared software or data and use them as a seed for a
next iteration.

\begin{quote}
\emph{Download the key papers and set them aside.} \emph{Then go to the
websites of the key researchers and see where they've published
recently.} \emph{That will help you identify the top conferences in that
field because the best researchers usually publish in the top
conferences.}
\end{quote}

Also check where they publish their code and data. It will give you an
idea where this community interacts online and can even lead you to
research compendia under development.

\begin{quote}
\emph{The third step is to go to the website for these top conferences
and look through their recent proceedings.} \emph{A quick scan will
usually identify recent high-quality related work.} \emph{These papers,
along with the ones you set aside earlier, constitute the first version
of your survey.} \emph{Make two passes through these papers.} \emph{If
they all cite a key paper that you did not find earlier, obtain and read
it, iterating as necessary.} (Keshav 2016)
\end{quote}

If a majority cites or uses a key software, technology, or dataset, then
evaluate it and include it in the next iteration.

\hypertarget{related-work}{%
\section*{4. Related work}\label{related-work}}
\addcontentsline{toc}{section}{4. Related work}

\begin{quote}
\emph{If you are reading a paper to do a review, you should also read
Timothy Roscoe's paper on ``Writing reviews for systems conferences''
(Roscoe 2007).} \emph{If you're planning to write a technical paper, you
should refer both to Henning Schulzrinne's comprehensive web site
(Schulzrinne n.d.) and George Whitesides's excellent overview of the
process (Whitesides 2004).} \emph{Finally, Simon Peyton Jones has a
website that covers the entire spectrum of research skills (Peyton Jones
n.d.).} \emph{Iain H. McLean of Psychology, Inc.~has put together a
downloadable `review matrix' that simplifies paper reviewing using the
three-pass approach for papers in experimental psychology (McLean 2012),
which can probably be used, with minor modifications, for papers in
other areas.} (Keshav 2016)
\end{quote}

We are working on an extended version of this matrix to provide space
for notes about software, data, results of the reproduction, and
application of the methods. See the corresponding repository issue for
details and provide your feedback:
\url{https://github.com/nuest/how-to-read-a-research-compendium/issues/2}

If you are reviewing a research compendium, a more detailed checklist is
given in the ``rOpenSci Analysis Best Pratice Guidelines'' (rOpenSci
2017), which are partially even automated for R-based research compendia
(DeCicco et al. 2018), and the Journal of Open Research Software's
guidelines for reviewing research software (JORS Editorial Team 2018).

\hypertarget{acknowledgements}{%
\section*{5. Acknowledgements}\label{acknowledgements}}
\addcontentsline{toc}{section}{5. Acknowledgements}

\begin{quote}
\emph{The first version of this document was drafted by my students:
Hossein Falaki, Earl Oliver, and Sumair Ur Rahman.} \emph{My thanks to
them.} \emph{I also benefited from Christophe Diot's perceptive comments
and Nicole Keshav's eagle-eyed copy-editing.} \emph{I would like to make
this a living document, updating it as I receive comments.} \emph{Please
take a moment to email me any comments or suggestions for improvement.}
\emph{Thanks to encouraging feedback from many correspondents over the
years.} (Keshav 2016)
\end{quote}

In the spirit of the original paper, we would like to make this a living
document and invite readers to provide comments or suggestions for
improvement via email, as part of this preprint, or on the GitHub
repository:
\url{https://github.com/nuest/how-to-read-a-research-compendium}. The
repository also includes open questions and is where the paper's authors
openly discuss.

\hypertarget{references}{%
\section*{References}\label{references}}
\addcontentsline{toc}{section}{References}

\hypertarget{refs}{}
\leavevmode\hypertarget{ref-barba_terminologies_2018}{}%
Barba, Lorena A. 2018. ``Terminologies for Reproducible Research.''
\emph{arXiv:1802.03311 {[}Cs{]}}, February.
\url{http://arxiv.org/abs/1802.03311}.

\leavevmode\hypertarget{ref-barnes_publish_2010}{}%
Barnes, Nick. 2010. ``Publish Your Computer Code: It Is Good Enough.''
\emph{Nature News} 467 (7317):753--53.
\url{https://doi.org/10.1038/467753a}.

\leavevmode\hypertarget{ref-brinckman_computing_2018}{}%
Brinckman, Adam, Kyle Chard, Niall Gaffney, Mihael Hategan, Matthew B.
Jones, Kacper Kowalik, Sivakumar Kulasekaran, et al. 2018. ``Computing
Environments for Reproducibility: Capturing the `Whole Tale'.''
\emph{Future Generation Computer Systems}, February.
\url{https://doi.org/10.1016/j.future.2017.12.029}.

\leavevmode\hypertarget{ref-bryan_naming_2015}{}%
Bryan, Jenny. 2015. ``Naming Things.'' \emph{Speaker Deck}.
\url{https://speakerdeck.com/jennybc/how-to-name-files}.

\leavevmode\hypertarget{ref-chirigati_reprozip:_2016}{}%
Chirigati, Fernando, Rémi Rampin, Dennis Shasha, and Juliana Freire.
2016. ``ReproZip: Computational Reproducibility with Ease.'' In
\emph{Proceedings of the 2016 International Conference on Management of
Data}, 2085--8. SIGMOD '16. New York, NY, USA: ACM.
\url{https://doi.org/10.1145/2882903.2899401}.

\leavevmode\hypertarget{ref-claerbout_seventeen_1994}{}%
Claerbout, Jon. 1994. ``Seventeen Years of Super Computing.''
\url{http://sepwww.stanford.edu/sep/jon/nrc.html}.

\leavevmode\hypertarget{ref-decicco_checkers:_2018}{}%
DeCicco, Laura, Noam Ross, Alice Daish, Molly Lewis, Nistara Randhawa,
Carl Boettiger, Nils Gehlenborg, Jennifer Thompson, and Nicholas
Tierney. 2018. ``Checkers: Automated Checking of Best Practices for
Research Compendia.'' rOpenSci Labs.
\url{https://github.com/ropenscilabs/checkers}.

\leavevmode\hypertarget{ref-gentleman_statistical_2007}{}%
Gentleman, Robert, and Duncan Temple Lang. 2007. ``Statistical Analyses
and Reproducible Research.'' \emph{Journal of Computational and
Graphical Statistics} 16 (1):1--23.
\url{https://doi.org/10.1198/106186007X178663}.

\leavevmode\hypertarget{ref-green_computational_2018}{}%
Green, Seth Ariel, and April Clyburne-Sherin. 2018. ``Computational
Reproducibility via Containers in Social Psychology.'' \emph{PsyArXiv},
February. \url{https://doi.org/10.17605/OSF.IO/MF82T}.

\leavevmode\hypertarget{ref-ioannidis_how_2014}{}%
Ioannidis, John P. A. 2014. ``How to Make More Published Research
True.'' \emph{PLOS Medicine} 11 (10):e1001747.
\url{https://doi.org/10.1371/journal.pmed.1001747}.

\leavevmode\hypertarget{ref-jimenez_popper_2017}{}%
Jimenez, I., M. Sevilla, N. Watkins, C. Maltzahn, J. Lofstead, K.
Mohror, A. Arpaci-Dusseau, and R. Arpaci-Dusseau. 2017. ``The Popper
Convention: Making Reproducible Systems Evaluation Practical.'' In
\emph{2017 IEEE International Parallel and Distributed Processing
Symposium Workshops (IPDPSW)}, 1561--70.
\url{https://doi.org/10.1109/IPDPSW.2017.157}.

\leavevmode\hypertarget{ref-jors_editorial_team_journal_2018}{}%
JORS Editorial Team. 2018. ``Journal of Open Research Software -
Editorial Policies, Peer Review Process.''
\url{http://openresearchsoftware.metajnl.com/about/editorialpolicies/}.

\leavevmode\hypertarget{ref-kahneman_new_2014}{}%
Kahneman, Daniel. 2014. ``A New Etiquette for Replication.'' \emph{Soc.
Psychol.} 45 (4):310.

\leavevmode\hypertarget{ref-keshav_how_2007}{}%
Keshav, S. 2007. ``How to Read a Paper.'' \emph{SIGCOMM Comput. Commun.
Rev.} 37 (3):83--84. \url{https://doi.org/10.1145/1273445.1273458}.

\leavevmode\hypertarget{ref-keshav_how_2016}{}%
---------. 2016. ``How to Read a Paper.'' Manuscript. Waterloo, ON,
Canada.
\url{http://blizzard.cs.uwaterloo.ca/keshav/home/Papers/data/07/paper-reading.pdf}.

\leavevmode\hypertarget{ref-kluyver_jupyter_2016}{}%
Kluyver, Thomas, Benjamin Ragan-Kelley, Fernando Pérez, Brian Granger,
Matthias Bussonier, Jonathan Frederic, Kyle Kelley, et al. 2016.
``Jupyter Notebooks - a Publishing Format for Reproducible Computational
Workflows.'' \emph{Positioning and Power in Academic Publishing:
Players, Agents and Agendas}, 87--90.
\url{https://doi.org/10.3233/978-1-61499-649-1-87}.

\leavevmode\hypertarget{ref-knuth_literate_1984}{}%
Knuth, Donald E. 1984. ``Literate Programming.'' \emph{Comput. J.} 27
(2):97--111. \url{https://doi.org/10.1093/comjnl/27.2.97}.

\leavevmode\hypertarget{ref-konkol_-depth_2018}{}%
Konkol, Markus, and Christian Kray. 2018. ``In-Depth Examination of
Spatio-Temporal Figures in Open Reproducible Research.''
\emph{EarthArXiv}, April. \url{https://doi.org/10.17605/OSF.IO/Q53M8}.

\leavevmode\hypertarget{ref-konkol_state_2018}{}%
Konkol, Markus, Christian Kray, and Max Pfeiffer. 2018. ``The State of
Reproducibility in the Computational Geosciences.''
\url{https://doi.org/10.17605/osf.io/kzu8e}.

\leavevmode\hypertarget{ref-marwick_packaging_2018}{}%
Marwick, Ben, Carl Boettiger, and Lincoln Mullen. 2018. ``Packaging Data
Analytical Work Reproducibly Using R (and Friends).'' \emph{The American
Statistician} 72 (1):80--88.
\url{https://doi.org/10.1080/00031305.2017.1375986}.

\leavevmode\hypertarget{ref-mclean_literature_2012}{}%
McLean, Iain H. 2012. \emph{Literature Review Matrix}.
\url{http://archive.org/details/LiteratureReviewMatrix}.

\leavevmode\hypertarget{ref-nosek_promoting_2015}{}%
Nosek, B. A., G. Alter, G. C. Banks, D. Borsboom, S. D. Bowman, S. J.
Breckler, S. Buck, et al. 2015. ``Promoting an Open Research Culture.''
\emph{Science} 348 (6242):1422--5.
\url{https://doi.org/10.1126/science.aab2374}.

\leavevmode\hypertarget{ref-nust_opening_2017}{}%
Nüst, Daniel, Markus Konkol, Edzer Pebesma, Christian Kray, Marc
Schutzeichel, Holger Przibytzin, and Jörg Lorenz. 2017. ``Opening the
Publication Process with Executable Research Compendia.'' \emph{D-Lib
Magazine} 23 (1/2). \url{https://doi.org/10.1045/january2017-nuest}.

\leavevmode\hypertarget{ref-pebesma_earth_2013}{}%
Pebesma, Edzer. 2013. ``Earth and Planetary Innovation Challenge (EPIC)
Submission "One-Click-Reproduce".''
\url{http://pebesma.staff.ifgi.de/epic.pdf}.

\leavevmode\hypertarget{ref-peyton_jones_simon_nodate}{}%
Peyton Jones, Simon. n.d. ``Simon Peyton Jones at Microsoft Research.''
\emph{Simon Peyton Jones at Microsoft Research}. Accessed May 25, 2018.
\url{https://www.microsoft.com/en-us/research/people/simonpj/}.

\leavevmode\hypertarget{ref-prlic_ten_2012}{}%
Prlić, Andreas, and James B. Procter. 2012. ``Ten Simple Rules for the
Open Development of Scientific Software.'' \emph{PLOS Comput Biol} 8
(12):e1002802. \url{https://doi.org/10.1371/journal.pcbi.1002802}.

\leavevmode\hypertarget{ref-ropensci_ropensci_2017}{}%
rOpenSci. 2017. ``rOpenSci Analysis Guide (Unconf 2017).'' \emph{Google
Docs}.
\url{https://docs.google.com/document/d/1OYcWJUk-MiM2C1TIHB1Rn6rXoF5fHwRX-7_C12Blx8g/edit?usp=embed_facebook}.

\leavevmode\hypertarget{ref-roscoe_writing_2007}{}%
Roscoe, Timothy. 2007. ``Writing Reviews for Systems Conferences,''
March, 6.
\url{https://people.inf.ethz.ch/troscoe/pubs/review-writing.pdf}.

\leavevmode\hypertarget{ref-sandve_ten_2013}{}%
Sandve, Geir Kjetil, Anton Nekrutenko, James Taylor, and Eivind Hovig.
2013. ``Ten Simple Rules for Reproducible Computational Research.''
\emph{PLoS Comput Biol} 9 (10):e1003285.
\url{https://doi.org/10.1371/journal.pcbi.1003285}.

\leavevmode\hypertarget{ref-schulzrinne_writing_nodate}{}%
Schulzrinne, Henning. n.d. ``Writing Systems and Networking Articles.''
Accessed May 25, 2018.
\url{https://www.cs.columbia.edu/~hgs/etc/writing-style.html}.

\leavevmode\hypertarget{ref-stack_overflow_how_2018}{}%
Stack Overflow. 2018. ``How to Create a Minimal, Complete, and
Verifiable Example.'' \emph{Stack Overflow}.
\url{https://stackoverflow.com/help/mcve}.

\leavevmode\hypertarget{ref-stodden_legal_2009}{}%
Stodden, Victoria. 2009. ``The Legal Framework for Reproducible
Scientific Research: Licensing and Copyright.'' \emph{Computing in
Science \& Engineering} 11 (1):35--40.
\url{https://doi.org/10.1109/MCSE.2009.19}.

\leavevmode\hypertarget{ref-stodden_best_2014}{}%
Stodden, Victoria, and Sheila Miguez. 2014. ``Best Practices for
Computational Science: Software Infrastructure and Environments for
Reproducible and Extensible Research.'' \emph{Journal of Open Research
Software} 2 (1). \url{https://doi.org/10.5334/jors.ay}.

\leavevmode\hypertarget{ref-stodden_researchcompendia.org:_2015}{}%
Stodden, Victoria, Sheila Miguez, and Jennifer Seiler. 2015.
``ResearchCompendia.Org: Cyberinfrastructure for Reproducibility and
Collaboration in Computational Science.'' \emph{Computing in Science \&
Engineering} 17 (1):12--19. \url{https://doi.org/10.1109/MCSE.2015.18}.

\leavevmode\hypertarget{ref-taschuk_ten_2017}{}%
Taschuk, Morgan, and Greg Wilson. 2017. ``Ten Simple Rules for Making
Research Software More Robust.'' \emph{PLOS Computational Biology} 13
(4):e1005412. \url{https://doi.org/10.1371/journal.pcbi.1005412}.

\leavevmode\hypertarget{ref-tatham_how_nodate}{}%
Tatham, Simon. n.d. ``How to Report Bugs Effectively.'' Accessed June
11, 2018. \url{https://www.chiark.greenend.org.uk/~sgtatham/bugs.html}.

\leavevmode\hypertarget{ref-vandevalle_code_2012}{}%
Vandevalle, Patrick. 2012. ``Code Sharing Is Associated with Research
Impact in Image Processing.'' \emph{Computing in Science \& Engineering}
Reproducible Research for Scientific Computing (July):42--47.

\leavevmode\hypertarget{ref-whitesides_whitesides_2004}{}%
Whitesides, G. M. 2004. ``Whitesides' Group: Writing a Paper.''
\emph{Advanced Materials} 16 (15):1375--7.
\url{https://doi.org/10.1002/adma.200400767}.

\leavevmode\hypertarget{ref-wilson_good_2017}{}%
Wilson, Greg, Jennifer Bryan, Karen Cranston, Justin Kitzes, Lex
Nederbragt, and Tracy K. Teal. 2017. ``Good Enough Practices in
Scientific Computing.'' \emph{PLOS Computational Biology} 13
(6):e1005510. \url{https://doi.org/10.1371/journal.pcbi.1005510}.

\end{document}